\documentclass[twocolumn,aps,prl,showpacs,floatfix]{revtex4}
\usepackage{amsmath}
\usepackage{hyperref}
\usepackage{latexsym}
\usepackage{color}
\usepackage{epsfig}
\usepackage{amsmath}
\usepackage{amssymb}
\usepackage{dsfont}
\def\beq{\begin{equation}}
\def\eeq{\end{equation}}
\def\bea{\begin{eqnarray}}
\def\eea{\end{eqnarray}}

\newcommand{\CP}{{\mathcal P}}

\newcommand{\CA}{{\mathcal A}}
\def\ep{\epsilon}

\def\ri{{\rm{i}}}

\begin{document}
%%%%%%%%%%%%%%%%%%%%%%%%%%%%%%%%%%%%%%%%%%%%
%%%%%%%%%%%%%%%%%%%%%%%%%%%%%%%%%%%%%%%%%%%%
\title{On the Integrability  \\ of the Discrete Nonlinear Schr\"odinger Equation}
%%%%%%%%%%%%%%%%%%%%%%%%%%%%%%%%%%%%%%%%%%%%
%%%%%%%%%%%%%%%%%%%%%%%%%%%%%%%%%%%%%%%%%%%%
\author{Decio Levi}
\email{levi@roma3.infn.it}
\affiliation{Dipartimento di Ingegneria Elettronica, Universit\`a degli Studi Roma Tre and \\Sezione INFN Roma Tre,
Via della Vasca Navale 84, 00146 Roma, Italy}

\author{Matteo Petrera}
\email{petrera@fis.uniroma3.it}
\author{Christian Scimiterna}
\email{scimiterna@fis.uniroma3.it}
\affiliation{Dipartimento di Fisica, Universit\`a degli Studi Roma Tre and \\ Sezione INFN Roma Tre,
Via della Vasca Navale 84, 00146 Roma, Italy}
%%%%%%%%%%%%%%%%%%%%%%%%%%%%%%%%%%%%%%%%%%%%
%%%%%%%%%%%%%%%%%%%%%%%%%%%%%%%%%%%%%%%%%%%%
\begin{abstract}
In this letter we present an analytic evidence of the non-integrability of the
discrete nonlinear 
Schr\"odinger equation,
a well-known discrete evolution equation which has been obtained 
in various contexts of physics and biology. We use a reductive
perturbation
technique to show an obstruction to its  integrability.

\end{abstract}
%%%%%%%%%%%%%%%%%%%%%%%%%%%%%%%%%%%%%%%%%%%%
%%%%%%%%%%%%%%%%%%%%%%%%%%%%%%%%%%%%%%%%%%%%

\date{\today}
%%%%%%%%%%%%%%%%%%%%%%%%%%%%%%%%%%%%%%%%%%%%
%%%%%%%%%%%%%%%%%%%%%%%%%%%%%%%%%%%%%%%%%%%%
\maketitle
%%%%%%%%%%%%%%%%%%%%%%%%%%%%%%%%%%%%%%%%%%%%
%%%%%%%%%%%%%%%%%%%%%%%%%%%%%%%%%%%%%%%%%%%%

%%%%%%%%%%%%%%%%%%%%%%%%%%%%%%%%%%%%%%%%%%%%
%%%%%%%%%%%%%%%%%%%%%%%%%%%%%%%%%%%%%%%%%%%%
\section{Introduction}
%%%%%%%%%%%%%%%%%%%%%%%%%%%%%%%%%%%%%%%%%%%%
%%%%%%%%%%%%%%%%%%%%%%%%%%%%%%%%%%%%%%%%%%%%

The  nonlinear Schr\"odinger (NLS) equation 
\beq
 \ri \partial_t f + \partial_{xx} f = \sigma |f|^2 f,\qquad f=f(x,t), \qquad
 \sigma =\pm 1 \,,\label{e0}
\eeq
 is a {\it universal}  
nonlinear integrable partial differential equation (PDE) for models with weak
nonlinear effects \cite{ca}. 
It has been central for almost
fourty years in a large variety of areas in sciences and 
it appears in many physical contexts, see for instance \cite{AS,APT,SS,AC,H}.
From the integrability of this PDE it follows the existence of
infinitely many symmetries and conservation laws, and the possibility 
of solving its Cauchy problem,  once the
initial data are prescribed. In correspondence with its symmetries one 
finds an infinite number of exact solutions, the solitons,
which, up to a phase, emerge unperturbed from
the interaction among themselves. 

Many  physical and biological applications involve lattice systems. In literature, one may find
a few discrete forms of the NLS equation.
The most relevant  one is the discrete NLS (DNLS) equation
\beq
\ri\partial_{t}f_{n}+\frac{1}{2 h^2} (f_{n+1}-2f_{n}+f_{n-1})
= \sigma |f_{n}|^2 f_{n}\, , \label{ee1}
\eeq
where $n \in \mathbb Z$ is a discrete index, $t$ is a real variable, $f_n$ is a complex function
and $h$ is a real parameter
related to the space-discretization. Its 
continuous limit ($h \rightarrow 0$, $n \rightarrow \infty$, $x=nh$ finite) 
goes into the integrable NLS equation (\ref{e0}).
Eq.  (\ref{ee1})
is one of the most studied lattice models (see for instance
\cite{APT,AF,DN, Surv_Tsir,Surv_Flach,Eilbeck,Bose} and references therein). 
Among  the many recent
applications of Eq. (\ref{ee1}) let us just mention the theory of
Bose-Einstein condensates in optical lattices \cite{Bose} and semiconductors \cite{Surv_Tsir}.
The DNLS equation possesses exact  discrete breathers 
solutions,  where the bumps are spatially localized and
periodic in time  \cite{Surv_Flach,abk}. However, just a few number of conserved quantities is known and thus
the DNLS equation (\ref{ee1}) is supposed to be non-integrable \cite{Ablowitz_Francoise}. Numerical schemes have been used to
exhibit its chaotic behavior \cite{AOT}. As far as we know no proof of its non-integrability is known and a few articles can be found on this subject \cite{CK}.

An integrable discretization of the NLS equation has been found
by Ablowitz and Ladik \cite{AL}. It reads
\bea
&& \ri\partial_{t}f_{n}+\frac{1}{2 h^2} (f_{n+1}-2f_{n}+f_{n-1}) = \nonumber \\
&& = \frac{\sigma}{2} |f_{n}|^2 (f_{n+1}+f_{n-1})\, . \label{e2}
\eea
Eq. (\ref{e2}) has an infinite number of explicit standing as well as travelling soliton solutions \cite{APT}.  In the limit
as $h \rightarrow 0$, also Eq. (\ref{e2}) goes into the NLS equation (\ref{e0}).
 Eqs. (\ref{ee1}) and (\ref{e2})  exhibit
very different responses to the same initial data. In the case of an infinite lattice with
rapidly decaying boundary conditions, Eq. (\ref{e2}) has soliton solutions while Eq. (\ref{ee1}) does not \cite{AOT,KK}.

Multiscale perturbation techniques \cite{TN}
have proved to be
important tools for finding approximate solutions to many  physical
problems by reducing a given PDE
to a simpler equation, which can be integrable 
\cite{ca}. 
These multiscale expansions are structurally strong and 
can be applied to both integrable and non-integrable systems.
Zakharov and Kuznetsov \cite{ZK} have shown that,
starting from an integrable PDE and performing a proper multiscale
expansion, one may obtain other integrable systems. In particular, they showed
that the slow-varying amplitudes of a dispersive wave solution of Eq. (\ref{e0}) satisfies the  Korteweg-de Vries (KdV) equation (\ref{e0}) and
{\it viceversa}. Calogero et al. \cite{ce} have used the multiscale perturbation technique as a tool to give necessary conditions for the integrability of large classes of PDE's both in 1+1 and 2+1 dimensions. In particular the non-integrability of the resulting multiscale reduction is a consequence of the non-integrability of the ancestor system. 
Multiscale techniques have been used also in the papers \cite{DMS,DP,MK} to prove integrability 
of several PDE's.

Some attempts to
extend this approach  to discrete equations have been
proposed \cite{Schoo,Ag,lm,LevHer,levi,lp,HLPS,HLPS2,HLPS3}. In
\cite{lm,levi,lp} one can find a 
multiscale expansion technique on
the lattice which, starting from dispersive integrable $\mathbb{Z}^2$-lattice equations,
provides other $\mathbb{Z}^2$-lattice equations.  
To do so one had to introduce a slow-varying condition on the amplitudes by requiring
\bea \label{55}
(\Delta_n)^{p+1} f_n = 0\, ,
\eea
$p$ being a positive integer and $\Delta_n f_n = f_{n+1}-f_n$. As a consequence, the resulting reduced
equation
turned out to be non-integrable even if the ancestor equation was integrable. 
However, as shown in \cite{HLPS}, 
if $p=\infty$ the reduced equations become  formally continuous and
their 
integrability may be preserved by the discrete reductive perturbation procedure. 
In this way  the multiscale expansions easily fit with
both difference-difference and differential-difference equations.
The illustrative example considered in 
\cite{HLPS} has been the lattice potential KdV equation, a dispersive
nonlinear $\mathbb{Z}^2$-lattice 
equation obtained from the superposition formula for the soliton solutions of
the KdV equation.
There one performed the multiscale
expansion of the weak plane wave solutions 
of the discrete dispersive linear system. A proper representation of the discrete shift operators in
terms of differential ones,
provides the integrable NLS equation (\ref{e0}) as
the lowest order secularity condition from
the multiscale expansion. 
Further examples have been considered in \cite{HLPS3}. They confirmed a
discrete analog of the Zakharov-Kuznetsov's claim \cite{ZK}:
{\it ``if a nonlinear dispersive discrete equation is integrable
then its lowest order multiscale reduction is an integrable NLS
equation''}. 

In the present letter we consider the multiscale
perturbation analysis of Eq. (\ref{ee1}). Multiscale analysis of the DNLS
equation (\ref{ee1}) has been already considered in \cite{Ag}, giving a
differential-difference system which does not  fulfill any integrability
criterium. However this result
did not prove the non-integrability of the DNLS equation, as similar results
have been obtained in the case of the integrable equation
(\ref{e2}) \cite {Ag,LevHer}. Here, extending to lattice equations the approach 
 used in \cite{ZK,DMS,DP,MK}
and 
computing the higher order terms in the reductive perturbation 
expansion we are able to provide an {\it analytical} evidence of  the
non-integrability of Eq. (\ref{ee1}). In fact, even if its lowest order multiscale reduction
is an integrable KdV-type equation, the higher orders reductions
exhibit non-integrable behaviors.

%%%%%%%%%%%%%%%%%%%%%%%%%%%%%%%%%%%%%%%%%%%%
%%%%%%%%%%%%%%%%%%%%%%%%%%%%%%%%%%%%%%%%%%%%
\section{Discrete multiscale analysis of the DNLS equation}
%%%%%%%%%%%%%%%%%%%%%%%%%%%%%%%%%%%%%%%%%%%%
%%%%%%%%%%%%%%%%%%%%%%%%%%%%%%%%%%%%%%%%%%%%

By the position $f_n(t)= [\nu_n(t)]^{1/2} {\rm{exp}} [\ri \phi_{n}(t) ]$
the DNLS equation (\ref{ee1})
 may be written as the following system of real
differential-difference equations:
\bea
&& \!\!\!\!\!\!\!\! \!\!\!\!\partial_{t}\nu_{n}+\frac{1}
{h^2}\left( \sqrt{\alpha_n^{+}}\sin\beta_n^{+}+\sqrt{\alpha_n^{-}} \sin\beta_n^{-}\right)=0\,,\label{nh1}\\
&& \!\!\!\!\!\!\!\! \!\!\!\!\partial_{t}\phi_{n}+\frac{1} {h^2}-\frac{1}
{2h^2}\left(\sqrt{\gamma_n^{+}}\cos\beta_n^{+}+\sqrt{\gamma_n^{-}}\cos\beta_n^{-}\right)+  \label{nh2}  \\
&& \;+ \, \sigma\nu_{n}=0\,, \nonumber
\eea
where $\alpha^{\pm}_n(t)=\nu_{n}\nu_{n\pm 1}$, 
$\beta_n^{\pm}(t)=\phi_{n\pm 1}-\phi_{n}$, and
$\gamma^{\pm}_n(t)=\nu_{n}^{-1}\nu_{n\pm 1}$.
By analogy with the continuous case, see \cite{ZK}, we expand the real fields $\nu_n(t)$
and $\phi_n(t)$ around the constant solution $f_n(t)=\exp{(-\ri\sigma t)}$ of the DNLS:
\bea
&&\nu_{n}(t)=1+\sum_{i=1}^{\infty}
\ep^{2i} \,\nu^{(i)}(\kappa, \{t_{m}\}_{m \geq 1})\, ,\label{exp1}\\
&&\phi_{n}(t)=-\sigma t+\sum_{i=1}^{\infty}
\ep^{2i-1} \,\phi^{(i)}(\kappa, \{t_{m}\}_{m \geq 1})\, ,\label{exp2}
\eea
where $\ep$, with $0 \leq |\ep| \ll 1$, is the perturbation parameter.
The fields $\nu^{(j)}$ and 
$\phi^{(j)}$ in Eqs. (\ref{exp1}-\ref{exp2}) depend on the slow-space variable
$\kappa=  \ep \zeta n$, $\zeta \in \mathbb R$, and the slow-time variables $  t_m = \ep^{2m-1}t$, $m \geq 1$. The free parameter
$\zeta$  can be fixed so as to obtain the proper continuous limit. 

Given a function $u_n(t) = v(\kappa, \{t_{m}\}_{m \geq 1})$ we expand  $u_{n \pm
  1}(t)$ and $\partial_t u_n(t)$ in terms of the slow 
 variables $\kappa$ and $\{t_{m}\}_{m\geq 1}$. Introducing the shift operator $T_n$ such that $T_n^{\pm} u_n = u_{n \pm 1}$ we have 
\bea
\!\!\!\!\!  u_{n \pm 1}(t) &=&   \left(T_{\kappa}^{\pm}\right)^{\epsilon \zeta} v(\kappa, \{t_{m}\}_{m \geq 1}) = \nonumber \\
&=&  
\sum_{i=0}^\infty \epsilon^i \CA_i^\pm v(\kappa, \{t_{m}\}_{m \geq 1})\, ,  \label{pdio}
\\ \label{pdioa}  
\!\!\!\!\! \partial_t  u_{n}(t) &=& \sum_{i=1}^{\infty} \epsilon^{2i-1}\partial_{t_i}
v(\kappa, \{t_{m}\}_{m \geq 1})\, ,
\eea
where, defining  $\Delta_{\kappa}^i= (T_{\kappa}-1)^i$ to be the $i$-th order difference
operator, one get 
\beq
\CA_i^\pm = \frac{(\pm  \zeta \delta_\kappa)^i}{i!} \, , \quad 
\delta_{\kappa} =\sum_{i=1}^\infty
\frac{(-1)^{i-1}}{i}\Delta_{\kappa}^i\, . \label{mnb}
\eeq
If  $u_n$ is  
a slow-varying function of order $p$, see Eq. (\ref{55}),
we can truncate the infinite series in Eq. (\ref{mnb}).
 In such a case the $\delta_\kappa$-operators  reduce
to polynomials in the $\Delta_\kappa$-operators of order at most $p$. 
Hereafter we shall assume $p=\infty$ and the
$\delta_\kappa$-operators 
are formal differential operators.

Taking into account the expansions (\ref{exp1}-\ref{exp2}) and Eq. (\ref{pdio}) we have the following formulas for the
shifts of the functions $\nu_n(t)$ and $\phi_n(t)$:
\bea
&&\!\!\!\!\! \!\!\!\!\!\!\!\!\!\!  \nu_{n\pm
1}(t)=1+\sum_{j=2}^{+\infty}\sum_{i=1}^{\left[j/2\right]}\ep^j\CA^{\pm}_{j-2i}
\nu^{(i)}(\kappa, \{t_{m}\}_{m \geq 1})\, ,\label{f1}\\
&&\!\!\!\!\! \!\!\!\!\! \!\!\!\!\! \phi_{n\pm 1}(t)=-\sigma
t+  \nonumber \\
&& \; \; \, + \sum_{j=1}^{+\infty}\sum_{i=1}^{\left[(j+1)/2\right]}\ep^j
\CA^{\pm}_{j-2i+1}\phi^{(i)}(\kappa, \{t_{m}\}_{m \geq 1})\,  ,\label{f2}
\eea
where $[x]$ denotes the integer part of $x$.

Let us introduce the multiscale expansions (\ref{f1}-\ref{f2}), together with
Eq. (\ref{pdioa}), into
Eqs. (\ref{nh1}-\ref{nh2}) and 
require that these equations be satified at all orders in $\ep$. 

At the lowest non-trivial order $\ep^2$ one finds
$\nu^{(1)}=-\sigma \partial_{t_{1}}\phi^{(1)}$. From now on all results will
be presented just for the functions $\phi^{(i)}$. 

At order $\ep^3$ we get
$\left(\partial_{t_{1}}^2-c^2\delta_{\kappa}^2\right)\phi^{(1)}=0$, where $c=\pm \left( \zeta \sigma^{1/2}\right)/ h$. As $c$ has to be real
then $\sigma=1$; moreover we choose $\zeta=h$ so that $c$ remains finite as
$h\rightarrow 0$. Therefore the resulting equation at this order is satisfied by
$\phi^{(1)}=  \phi^{(1)} (\xi , \{t_m\}_{m \geq 2})$ with $\xi= \kappa - c t_1$.

At order  $\ep^5$,  the no-secular term condition implies  
$\left(\partial_{t_{1}}^2-c^2 \delta_{\kappa}^2\right)\phi^{(2)}=0$,
i.e. $\phi^{(2)}=  \phi^{(2)} (\xi , \{t_m\}_{m \geq 2})$. The evolution equation for 
$\phi^{(1)}$ w.r.t. the slow-time $t_2$ reads
\bea
&&\partial_{t_{2}}\phi^{(1)}= K_2\left(  \phi^{(1)}\right)\, ,\label{Caesar} \\
&& K_2\left(  \phi^{(1)}\right) = a \left[
 \partial_{\xi}^3\phi^{(1)}- \frac{3}
 {4a}\left(\partial_{\xi}\phi^{(1)}\right)^2 \right]\, ,\nonumber 
\eea
with $a= c(3-h^2)/24$. Eq. (\ref{Caesar}) is 
 a potential KdV equation. Therefore, if Eq. (\ref{ee1}) has to be integrable, then its
multiscale reductions should provide the integrable evolution equations ($j \geq 3$):
\bea \label{eq:hpkdv}
&&\!\!\!\!\! \!\!\!\!\!  \partial_{t_{j}}\phi^{(1)}=K_j(\phi^{(1)})= b_j \int^\xi dy \,   \mathcal L^{j-1}
\left[\partial_{y}^2 \phi^{(1)}\right] \,  ,
\\
&& \!\!\!\!\! \!\!\!\!\! \mathcal L \left[f(\xi)\right] = \partial_{\xi}^2 f(\xi)
-\frac{\partial_{\xi} \phi^{(1)}}{a} f(\xi) - \frac{\partial_{\xi}^2 \phi^{(1)}}{2a}
\int^\xi dy f(y) \, , \nonumber 
\eea
where $\mathcal L$ is the recursive operator associated with the KdV hierarchy
and the $b_j$'s are free coefficients to be fixed later.

We assign a formal degree to the $\xi$-derivatives of the functions 
$\phi^{(j)}$, $ {\rm{deg}} \left(  \partial_\xi^\ell \phi^{(j)}\right)= \ell+
2j-1$, $\ell \geq 0$, and  define  $\CP_n$  as the vector space spanned by the products with total degree $n$ of all 
derivatives $\partial_\xi^\ell \phi^{(j)}$. We denote
by $\CP_n^{(r)} \subset \CP_n$ the subspace spanned by those products of derivatives
$\partial_\xi^\ell \phi^{(j)}$ with
$j \leq r$. Similar vector spaces have been introduced in \cite{DP,DMS} in the case of the
multiscale analysis of PDE's.

After caring for secularities, the order $\ep^7$ yields 
$\phi^{(3)}=  \phi^{(3)} (\xi , \{t_m\}_{m \geq 2})$  and
the following non-homogeneous  evolution equation for the field $\phi^{(2)}$ w.r.t. the slow-time $t_2$, depending 
on $\phi^{(1)}$ and its derivatives:
\bea
&&\!\!\!\!\! \partial_{t_{2}}\phi^{(2)}- \frac{c (3-h^2)}{24}\partial_{\xi}^3\phi^{(2)}+
\frac{3} {2}\partial_{\xi}\phi^{(1)}\partial_{\xi}\phi^{(2)}= 
\label{Romulus}\\
&&\!\!\!\!\!  =-\partial_{t_{3}}\phi^{(1)} - \frac{5h^2-7}{64}\left(\partial_{\xi}^2\phi^{(1)}\right)^2 +
\frac{ch^2} {12}\left(\partial_{\xi}\phi^{(1)}\right)^3- \nonumber \\
&& \; - \, \frac{3h^2+1}{16}\partial_{\xi}\phi^{(1)}\partial_{\xi}^3\phi^{(1)}
- \frac{c (h^4-30h^2-15)}{1920} \partial_{\xi}^5\phi^{(1)}. \nonumber 
\eea 

Substituting Eq. (\ref{eq:hpkdv}) into Eq. (\ref{Romulus}) with $j=3$ and
fixing  $b_3=-c (h^4-30h^2-15)/1920$ in order
to remove residual secularities, Eq. (\ref{Romulus}) reduces to the following 
 evolution equation for the field $\phi^{(2)}$ w.r.t. the slow-time $t_2$:
\beq
\partial_{t_{2}}\phi^{(2)}-K_{2}^{\prime}\left(\phi^{(1)}\right)\phi^{(2)}=f^{(t_2)}\,
, \label{sd}
\eeq
where $K_{2}^{\prime}\left(\phi^{(1)}\right)\phi^{(2)}=(d / d \theta) K_2 (\phi^{(1)} + \theta \phi^{(2)}) |_{\theta=0}$ 
is the Fr\'echet derivative of $K_{2}\left(\phi^{(1)}\right)$ along the direction of $\phi^{(2)}$.
In Eq. (\ref{sd}) the forcing term $f^{(t_2)}$ is a well-defined element of $\CP_6^{(1)}$, namely
 a linear combination of three independent differential monomials, with coefficients that
are rational functions of $h$. 
At this same order, the integrability of Eq. (\ref{ee1}) implies the existence of the following  evolution equation
for the field $\phi^{(2)}$ w.r.t. the slow-time $t_3$:
\beq
\partial_{t_{3}}\phi^{(2)}-K_{3}^{\prime}\left(\phi^{(1)}\right)\phi^{(2)}=f^{(t_3)}\,
,\label{Pretestato} 
\eeq
where $f^{(t_3)}$ is an element of the space $\CP_8^{(1)}$, which if Eq. (\ref{ee1}) has to be integrable, must satisfy the 
compatibility condition $\left[\partial_{t_{3}}-K_{3}^{\prime}\left(\phi^{(1)}\right)\right]f^{(t_2)}=
\left[\partial_{t_{2}}-K_{2}^{\prime}\left(\phi^{(1)}\right)\right]f^{(t_3)}$. Such a condition allows one
to express the coefficients of the polynomial $f^{(t_3)}$ in terms of those of $f^{(t_2)}$, and does not impose any further constraint
on the coefficients of $f^{(t_2)}$.
As this condition is satisfied, eventual obstructions to the integrability of Eq. (\ref{ee1}) will
appear at higher perturbative orders.

Let us now consider the order $\ep^9$.
The resulting equations provide the evolution of the field $\phi^{(3)}$ w.r.t. the slow-time $t_{2}$. 
This is given by an integro-differential equation. Introducing the fields $\varphi^{(j)}=\partial_{\xi}\phi^{(j)}$, taking care of  secularities and taking into
account that $\phi^{(1)}$ evolves w.r.t. the slow-time $t_4$ according to Eq. (\ref{eq:hpkdv}),  we get
$\phi^{(4)}=  \phi^{(4)} (\xi , \{t_m\}_{m \geq 2})$ and
\beq
\partial_{t_{2}}\varphi^{(3)}-H_{2}^{\prime}\left(\varphi^{(1)}\right)\varphi^{(3)}=g^{(t_2)}\,
,\label{Angerona}
\eeq
where $H_{2}^{\prime}\left(\varphi^{(1)}\right)\varphi^{(3)}$ is 
the Fr\'echet derivative along $\varphi^{(3)}$ of the KdV flow 
$H_{2}\left(\varphi^{(1)}\right)=\partial_\xi K_2 \left(\phi^{(1)}\right)$.
Here $g^{(t_2)} \in \mathcal P_{9}^{(2)}$ is a linear combination of fourteen differential monomials, whose coefficients
are well-defined rational functions of $h$.
The evolution equation of $\varphi^{(3)}$ w.r.t. the slow-time $t_{3}$ takes the form
\beq
\partial_{t_{3}}\varphi^{(3)}-H_{3}^{\prime}\left(\varphi^{(1)}\right)\varphi^{(3)}=g^{(t_3)}\,
,\label{Tarqnas}
\eeq
where $H_{3}^{\prime}\left(\varphi^{(1)}\right)\varphi^{(3)}$ is the Fr\'echet derivative along $\varphi^{(3)}$ of the higher order KdV flow
$H_{3}\left(\varphi^{(1)}\right)= \partial_\xi K_3 \left(\phi^{(1)}\right)$. Here
$g^{(3)}_{t_3}$ is an element of the $31$-dimensional vector space $\mathcal P_{11}^{(2)}$ 
whose coefficients are determined by requiring the
compatibility condition 
\bea \label{56}
\left[\partial_{t_{3}}-H_{3}^{\prime}\left(\varphi^{(1)}\right)\right]g^{(t_2)}=
\left[\partial_{t_{2}}-H_{2}^{\prime}\left(\varphi^{(1)}\right)\right]g^{(t_3)}\, .
\eea
Eq. (\ref{56}) is a necessary condition for the integrability of Eq. (\ref{ee1}). 
In this case only nine out of the fourteen coefficients of $g^{(t_2)}$ are
independent. Thus we have five integrability conditions, whose explicit
expressions will be published elsewhere in a more detailed paper \cite{LPS}.
The obtained constraints on the polynomial $g^{(t_2)}$ are not
satisfied by the coefficients computed in Eq. (\ref{Angerona}). 
Then, this incompatibility implies that the DNLS equation (\ref{ee1}) cannot be 
 integrable.

%%%%%%%%%%%%%%%%%%%%%%%%%%%%%%%%%%%%%%%%%%%%
%%%%%%%%%%%%%%%%%%%%%%%%%%%%%%%%%%%%%%%%%%%%
\section{Concluding remarks} \label{concl}
%%%%%%%%%%%%%%%%%%%%%%%%%%%%%%%%%%%%%%%%%%%%
%%%%%%%%%%%%%%%%%%%%%%%%%%%%%%%%%%%%%%%%%%%%
By performing a discrete multiscale analysis, we have proven that the DNLS
equation
(\ref{ee1}) is not integrable. Although its lowest order reduction gives a KdV
equation, the higher order terms do not satisfy the required integrability
conditions. It is remarkable to note that a similar analysis performed
on the integrable discrete NLS equation (\ref{e2}) provides integrable
reductions \cite{LPS}.

Moreover, by using the expansions (\ref{exp1}-\ref{exp2}), 
our perturbation technique enables us to construct
approximate solutions of Eq. (\ref{ee1}) starting from the exact solutions
of the integrable Eq. (\ref{Caesar}).
However these solutions will break down in the
far-field region, due to non-integrability of Eq. (\ref{ee1}).

%%%%%%%%%%%%%%%%%%%%%%%%%%%%%%%%%%%%%%%%%%%%
%%%%%%%%%%%%%%%%%%%%%%%%%%%%%%%%%%%%%%%%%%%%
%\begin{acknowledgments}
%%%%%%%%%%%%%%%%%%%%%%%%%%%%%%%%%%%%%%%%%%%%
%%%%%%%%%%%%%%%%%%%%%%%%%%%%%%%%%%%%%%%%%%%%

%%%%%%%%%%%%%%%%%%%%%%%%%%%%%%%%%%%%%%%%%%%%
%%%%%%%%%%%%%%%%%%%%%%%%%%%%%%%%%%%%%%%%%%%%
%\end{acknowledgments}
%%%%%%%%%%%%%%%%%%%%%%%%%%%%%%%%%%%%%%%%%%%%
%%%%%%%%%%%%%%%%%%%%%%%%%%%%%%%%%%%%%%%%%%%%

%%%%%%%%%%%%%%%%%%%%%%%%%%%%%%%%%%%%%%%%%%%%
%%%%%%%%%%%%%%%%%%%%%%%%%%%%%%%%%%%%%%%%%%%%

%%%%%%%%%%%%%%%%%%%%%%%%%%%%%%%%%%%%%%%%%%%%
%%%%%%%%%%%%%%%%%%%%%%%%%%%%%%%%%%%%%%%%%%%%

\begin{thebibliography}{99}
%%%%%%%%%%%%%%%%%%%%%%%%%%%%%%%%%%%%%%%%%%%%
%%%%%%%%%%%%%%%%%%%%%%%%%%%%%%%%%%%%%%%%%%%%

\bibitem{ca} F. Calogero, 
in {\it What is integrability?}, edited by V.E. Zakharov (Springer, Berlin, 1991) 1. 

\bibitem{AS}
M.J. Ablowitz and H. Segur, 
{\it Solitons and the inverse scattering transform}
(SIAM, Philadelphia, 1981). 

\bibitem{AC} 
M.J. Ablowitz and P.A. Clarkson, 
{\it Solitons, nonlinear evolution equations and inverse scattering}
(Cambridge University Press, Cambridge, 1991).

\bibitem{APT}
M.J. Ablowitz, B. Prinari and A.D. Trubatch, 
{\it Discrete and continuous nonlinear Schr\"odinger systems}
(Cambridge University Press, Cambridge, 2004).


\bibitem{SS}
C. Sulem and P.L. Sulem, 
{\it The nonlinear Schr\"odinger equation}
(Springer-Verlag, New York, 1999).


\bibitem{H}
A. Hasegawa, 
{\it Solitons in optical communications} 
(Clarendon Press, Oxford, NY, 1995).


\bibitem{AF}
G. Assanto and A. Fratalocchi, Phys. Rev. A {\bf 76}, 042108 (2007).

\bibitem{DN}
A. Davydov and N. Kislukha, 
Phys. Stat. Sol. B {\bf 59}, 465 (1973).

\bibitem{Eilbeck} 
J.Ch. Eilbeck and M. Johansson, 
in {\it Localization and energy transfer in nonlinear
systems}, edited by L. V\'azquez, R.S. MacKay and M.P Zorzano
(World Scientific, Singapore, 2003), 44.

\bibitem{Surv_Flach}  
S. Flach and C.R. Willis, 
Phys. Reports {\bf 295}, 181 (1988).

\bibitem{Surv_Tsir}  
D. Hennig and G. Tsironis,
Phys. Reports {\bf 307}, 333 (1999).

\bibitem{Bose} 
F.Kh. Abdullaev, B.B. Baizakov, S.A. Darmanyan, V.V. Konotop and M. Salerno,
Phys. Rev. A {\bf 64}, 043606  (2001).

\bibitem{abk}
G.L. Alfimov, V.A. Brazhnyi and V.V. Konotop,
Phys. D {\bf 194}, 127  (2004) .

\bibitem{Ablowitz_Francoise}
M.J. Ablowitz and B. Prinari,
in {\it Encyclopedia of mathematical physics},
Vol. 3,  edited by J-P. Francoise, G.L. Naber and T.S. Tsun
(Academic Press/Elsevier Science, Oxford, 2006), 552.

\bibitem{AOT}
M.J. Ablowitz, Y. Ohta and A.D. Trubatch, 
Chaos  Sol. and Frac. {\bf 11}, 159 (2000).

\bibitem{CK} 
O. A. Chubykalo, V. V. Konotop and L. Vazquez,
Phys. Rev. B {\bf 47}, 7971 (1993).

\bibitem{AL}
M.J. Ablowitz and J.F. Ladik, 
Jour. Math. Phys. {\bf 17}, 1011 (1976).

\bibitem{KK}
T. Kapitula and P.G. Kevrekidis,
Nonlin.  {\bf 14}, 533 (2001).

\bibitem{TN}
T. Taniuti  and K. Nishihara,
{\it Nonlinear waves} 
(Pitman, Boston, 1983).


\bibitem{ZK}
V.E. Zakharov and E.A. Kuznetsov, 
Phys. D {\bf 18}, 455 (1986).

\bibitem{ce}
F. Calogero and W. Eckhaus,
Inv. Prob. {\bf 3}/2, L27  (1987);
F. Calogero and W. Eckhaus,
Inv. Prob. {\bf 3}/2, 229  (1987);
F. Calogero and W. Eckhaus,
Inv. Prob. {\bf 4}/1, 11 (1987);
F. Calogero, A. Degasperis and X-D. Ji, Jour. Math. Phys. {\bf 42}/6, 2635 (2001);
F. Calogero, A. Degasperis and X-D. Ji, Jour. Math. Phys. {\bf 41}/9, 6399 (2000);
F. Calogero and A. Maccari, in
{\it Inverse problems: an interdisciplinary study} (Academic Press, London, 1987) 463.

 \bibitem{DMS}
 A. Degasperis, S.V. Manakov and P.M. Santini,
 Phys. D {\bf 100}, 187 (1997).
 
\bibitem{DP}
A. Degasperis and M. Procesi,
in {\it Symmetry and perturbation theory, SPT98}, edited by A. Degasperis and G. Gaeta
(World Scientific, Singapore, 1999), 23; A. Degasperis, {\it Multiscale expansion and integrability
of dispersive wave equations}, Lectures given at the Euro Summer School ``What is integrability?", 13-24
August 2001, Isaac Newton Institute, Cambridge, UK.

\bibitem{MK}
Y. Kodama and A.V. Mikhailov, 
in {\it Algebraic aspects of integrable systems, 
Progr. Nonlinear Differential Equations Appl.}, 26, (BirkhŠuser Boston, Boston, MA, 1997) 173. 

\bibitem{Schoo}
S.W. Schoombie,
Jour. Comp. Phys. {\bf 101}, 55 (1992).


\bibitem{Ag}
M. Agrotis, S. Lafortune and P.G. Kevrekidis,
Discr. Cont. Dyn. Sist. {\bf 2005},  22 (2005).

\bibitem{HLPS} 
R. Hernandez Heredero, D. Levi, M. Petrera and C. Scimiterna, 
Jour. Phys. A {\bf 40}, F831 (2007).
 
 \bibitem{HLPS2} 
R. Hernandez Heredero, D. Levi, M. Petrera and C. Scimiterna, 
{\tt http://www.arxiv.org/abs/0709.3704}.

 \bibitem{HLPS3} 
R. Hernandez Heredero, D. Levi, M. Petrera and C. Scimiterna, 
{\tt http://www.arxiv.org/abs/0710.5299}.

\bibitem{lm}
J. Leon and M. Manna,
Jour. Phys. A {\bf 32}, 2845 (1999).

\bibitem{levi}
D. Levi,
Jour. Phys. A {\bf 38}, 7677 (2005).

\bibitem{LevHer}
D. Levi and R. Hernandez Heredero,
Jour. Nonlinear Math. Phys. {\bf 12}/1, 440 (2005).

\bibitem{lp}
D. Levi and M. Petrera,
 Jour. Math. Phys. {\bf 47}, 043509 (2006).
 


\bibitem{LPS} 
D. Levi, M. Petrera and C. Scimiterna, in preparation.




%%%%%%%%%%%%%%%%%%%%%%%%%%%%%%%%%%%%%%%%%%%%
%%%%%%%%%%%%%%%%%%%%%%%%%%%%%%%%%%%%%%%%%%%% 
\end{thebibliography}
\end{document}